%% file: main.tex
\documentclass{article}
\usepackage[utf8]{inputenc}

\title{Curatio et Innovatio}
\author{Roberto Rossi\\Business School, University of Edinburgh, UK\\roberto.rossi@ed.ac.uk}
\date{}
\usepackage[numbers]{natbib}
\usepackage{graphicx}
\usepackage[hyphens]{url}

\usepackage{listings}
\usepackage{copyleft}

\begin{document}
\maketitle

\begin{abstract}
The Middle Ages focused obsessively on the old; our era is totally absorbed with the new. In medio stat virtus. 
In this short note, I advocate a strategy that blends copyright and copyleft for disseminating research results in the sciences. 
I argue that such a blend may be beneficial in fields such as mathematics and computer science, that it may
facilitate the evolution and emergence of improved problem descriptions, whilst at the same time preserving author's rights, and easing researchers' work.
\end{abstract}

\section{Introduction}

In scientific enquiries, the researcher must acknowledge two essential elements. On the one hand, there is the Known; on the other, there is the Unknown.
The Known, is whatever has been investigated for years, decades, and perhaps centuries. People may have devoted their lives and efforts to structure a problem, demarcate it, dissect it, and come up with a polished description of the challenge, and of ways of dealing with it. 
The Unknown, conversely, is the unexplored. It is the realm of discovery, the force which keeps research going.

In this work, I argue that, while conducting research in the sciences, 
\begin{itemize}
    \item the Known and the Unknown should be approached in different ways, by leveraging two different tools: copyleft and copyright; and that
    \item dissemination of research results in the sciences should leverage a blend of these two strategies that deal with intellectual property.
\end{itemize}
 
The rest of this short note is structured as follows. In Section \ref{sec:known_unknown}, I survey different attitudes that existed in different ages towards the Known and the Unknown; in particular, the Middle Ages obsession with preservation of the Known, and the present time obsession with the discovery of the Unknown. In Section \ref{sec:copyleft_copyright} we introduce the concepts of copyright and copyleft. In Sections \ref{sec:curatio_innovatio}, I illustrate a possible new approach to dissemination of research results in the sciences, which attempts to leverage a blend of intellectual property protection strategies to strike a better balance between preservation of the Known, and discovery of the Unknown. This approach is operationalised in Section \ref{sec:practice}. In Section \ref{sec:conclusions}, I draw some concluding remarks.
 
\section{The Known and the Unknown}\label{sec:known_unknown}

The ``Known'' is the realm of the explored. It is that branch of knowledge that deals with problems that people have considered before, discussed, and perhaps solved, or left open. For instance, a known problem in mathematics is the Knapsack Problem \cite{10.2307/167356}, whose origin dates back to the early works of Dantzig. 
\begin{quote}
Medieval authorities tended to advocate a predominantly closed canon of knowledge which made dealing with previously unknown concepts particularly
difficult.\footnote{Foreign Knowledge – Medieval Attitudes towards the Unknown. In: H-Soz-Kult, 14.02.2018, \url{www.hsozkult.de/event/id/event-86250}.}
\end{quote}
``Curiositas,'' the process of seeking of new knowledge, was despised in the Middle Ages \cite{carruthers1998the}. 
\begin{quote}
    At the end of his discussion of a late medieval English guidebook to the Holy Land, Howard mentions the general medieval ambivalence toward curiosity. ``All this,'' he says, ``is curiositas --- the traveller's interest in what he sees and the reader's in what he hears. But it is exactly this `curiosity' that led pilgrims astray and put the pilgrimage in bad repute.'' \cite[p. 30]{atkinson1983}
\end{quote}
In a word, pilgrims ought not be concerned with discovering new lands and costumes; they had to focus on the spiritual aim of their journey. 
As a consequence of this established mindset, the Known ought to be preserved, commented upon, recombined (``varietas'' was prized in line with Ciceronian tenets of ancient rhetoric \cite{carruthers2013experience}), but seldom modified or expanded. Preserving the Known for future generations became a key aim and duty of medieval scholars, particularly in monastic settings. 

However, it is hard to preserve knowledge in its pristine state. When latin or greek manuscripts were transcribed by amanuenses, they were often annotated, and --- out of necessity --- modified. Changes were however minimal, unnoticeable. The action of the amanuenses was akin to an apprentice stonemason chiselling, akin to natural selection: unnoticeable in the grand scheme of things. And yet, that chiselling helped knowledge sail the waves of time, and evolve. Words were modified. Sentences were removed, inserted, or changed; and the original text therefore slowly morphed into a different one.

All this came to a dramatic change at the onset of the Renaissance. As discussed in \cite{wootton2015invention}, the change of mindset underpinning the Renaissance did not emerge overnight. The invention of the compass and of cartography --- its cognate discipline --- represent the prelude of the Age of Discovery. Sailors broke old established taboos and sailed through the ``Unknown'' to discover new land and riches. As a consequence of these discoveries, which directly contradicted established dogmas, people grew increasingly suspicious of the closed canon of knowledge that dominated the Middle Ages. These contradictions \cite{Kuhn:1970} catalysed a revolution in people's mind. The time was ripe for a dramatic change.

It is surprising to observe how key developments in human thought seem to have emerged almost synchronously. Within a time span of fifty years, a New World had been discovered (1492) by Christopher Columbus, and a ``Nova Scientia'' had been discussed by Nicola Tartaglia (1537) \cite{tartaglia2013nova}; and yet, Tartaglia's work was still ``conformist.'' Tartaglia's aim was to discuss a New Science, what we would now call dynamics, i.e. that branch of physics that deals with time-dependent physical matters. Akin to Euclid's work, which begins with the definition of a point, Tartaglia's work begins with the definition of an ``instant,'' that is a ``point in time;'' and then uses logic-deductive method
to derive results. However, Tartaglia did not have the means to go as far as Galileo went. He stopped short of Galileo's revolutionary claims, and  framed his New Science within an Aristotelian framework. Still, Tartaglia was able to obtain new results, and to chart the Unknown. In particular, he focused on a practical military problem, and showed at what angle a cannon should be fired, in order to achieve the longest possible shot \cite{citeulike:14287687}. It is this charting of the Unknown that made his science ``new;'' hence the title ``Nova Scientia,'' the New Science. A close inspection of Tartaglia's work reveals the influence he exerted on Galileo. 

\section{Copyright \& copyleft}\label{sec:copyleft_copyright}
Our age is clearly aligned with Tartaglia's and Galileo's mindset. Academics strive to produce new knowledge and to generate so-called ``impact.'' While doing so, they ``stand on the shoulders of giants.'' They build upon existing results to produce new knowledge. And yet, despite standing on the shoulders of giants, {\em a boulder blocks their view}: copyright. 

Copyright is a type of intellectual property that gives its owner the exclusive right to copy and distribute a creative work, usually for a limited time. The aim of copyright is to protect the original expression of an idea, but {\em not} the idea itself. Depending on the jurisdiction, there may exist limitations to copyright, e.g. fair use in the United States. Is it worth mentioning that the development of the concept of copyright is closely related to the invention of the printing press in the 15th and 16th centuries. At the onset, copyright came to exist in order to regulate what material could be printed. Eventually, it evolved into a set of laws that allow products of creative human activities to be preferentially exploited and thus, ideally, incentivised.

The concept of copyleft is perhaps less known than that of copyright. While copyright law gives software authors control over copying, distribution and modification of their works, the goal of copyleft is to give all users/viewers of the work the freedom to carry out all of these activities. There are four key freedoms, originally put forth in ``The Free Software Definition'' written by Richard Stallman and published by the Free Software Foundation (FSF) in 1986. The freedom to use the work; the freedom to study the work; the freedom to copy and share the work with others; the freedom to modify the work, and the freedom to distribute modified and therefore derivative works. The concept of copyleft originally emerged in the software development community. 

It is interesting to stress that the idea of ``free'' software originally did not refer to ``free of charge,'' but referred to ``free speech:'' a software free from undue constraints; a concept akin to the idea of Liberal Arts, which we find reflected in expressions such as Tartaglia's ``patet omnibus'' (open to everyone), referred to the New Science he was introducing. After gaining momentum in the software development community, the concept of Free Software evolved into that of Open Source content --- not just software --- as captured in the ``Open Source Definition'' originally published by the Open Source Initiative in the late nineties; and finally, the idea embraced an even broader scope by morphing, a decade later, into the notion of Open Knowledge,\footnote{\url{http://opendefinition.org/od/2.1/en/}} which promotes a robust commons in which anyone may participate, and interoperability is maximized. Dissemination of Open Knowledge is nowadays facilitated by a plethora of copyleft licenses, such as Creative Commons licenses.\footnote{\url{http://creativecommons.org}}

\section{Curatio et innovatio}\label{sec:curatio_innovatio}

When working on an existing problem, for example the Knapsack Problem, authors cannot reuse the original problem description verbatim --- for instance, in the case of the Knapsack Problem, the one presented in section ``The Knapsack Problem'' of \cite[][p. 273]{10.2307/167356}. In order to not infringe author's copyright, they must create a new description of the same problem, by paraphrasing the original text. This leads to a number of problems: for complex problems, some authors may misunderstand the problem description, and generate wrong, incomplete, or misleading paraphrases; other authors may develop correct but poor descriptions. Science is then caught in an endless cycle in which a problem description is constantly perturbed, constantly re-created afresh in every new paper published, and thus never reaches a steady state. Entropy, rather than stability, is sovereign.

I argue that creation of new knowledge --- for instance the discussion of a new algorithm to solve the Knapsack Problem --- and preservation of existing knowledge --- i.e. the ``best'' Knapsack Problem definition --- should be treated in different ways. In particular, copyright and copyleft should be used in concert while crafting research works.

More specifically, problem definitions should be disseminated under copyleft, e.g. Creative Commons Attribution (CC-BY), so that future authors and researchers may reuse them, and build upon them. I name this the ``curatio'' part of research, the preservation of knowledge, which is {\em completely overlooked by existing research practices}. The problem definition created by the original author should be initially published in a copyleft repository, rather than in a copyrighted work. This very same definition should then be reused in its original form by subsequent authors who aim to build upon the original research. Authors seeking to improve a problem definition, should themselves release their improved problem definition in copyleft form. Eventually, this strategy allows the ``best'' problem definition(s) to emerge spontaneously, by natural selection and popular vote: the best problem definitions being those that appear more frequently in published works, those definitions that are liked the most. 

Finally, new research results --- for example a new solution method for the Knapsack Problem --- new analyses, discussions, or other original findings may be published as usual, subject to copyright, to protect authors' and publishers rights. This is the ``innovatio'' part of research, which we are all familiar with.

\begin{figure}[ht]
    \centering
    \includegraphics[width=0.7\textwidth]{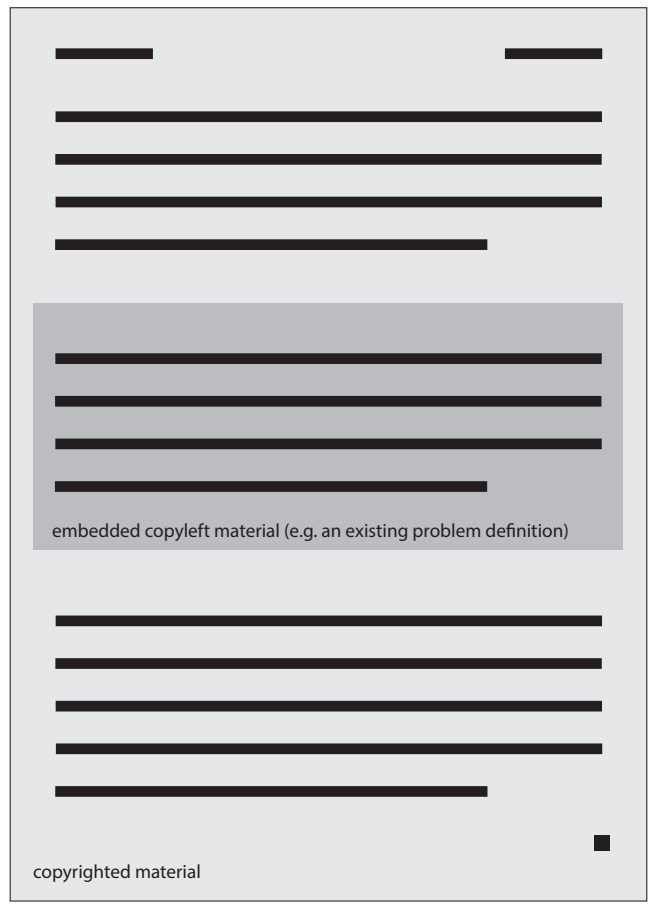}
    \caption{A sample page in a copyrighted work embedding copyleft materials.}
    \label{fig:sample_manuscript_page}
\end{figure}

How would then research be affected by this proposal? In essence, research will turn into a mix of ``curatio'' and of ``innovatio;'' of copyleft and copyright. 

Some researchers will focus on curating existing problems. They will focus on improving problem definitions, perhaps on public repositories,\footnote{e.g. GitHub, \url{https://www.github.com}; CSPLib, \url{https://www.csplib.org/}.} and on releasing copyleft versions of these definitions that will be reused in future research works. This will resemble what already happens in open source software development. These authors will be rewarded by seeing their problem definition, or a modified version of it, emerge as the definition of choice in the literature.

Other researchers, will focus on developing new approaches to tackle existing problems, or on developing new problems. While developing new approaches to tackle existing problems, they will publish as usual, in traditional journals, by leveraging and embedding the aforementioned copyleft materials (Fig. \ref{fig:sample_manuscript_page}). While developing new problems, they will take care to first release the problem definition in copyleft open source form, in public repositories, so that other researchers will be able to reuse the original text, embed it in their works, build upon it, modify it, and ideally improve it.

There are further advantages to this approach to disseminating science. The art of memory \cite{carruthers_2008} leverages patterns and repetitions.
Parataxis (juxtaposition) of traditional elements has been used since the dawn of mankind as a memory device. Its use is apparent in oral epic poetry, which required aoidoi to memorize entire poems. For instance, a device extensively used in Homer's works are ``fixed epithets,'' stereotyped descriptive phrases that can be leveraged as necessary to suit the demands of the metre: ``fleet-footed'' Achilles, ``wily''  Odysseus, or ``rosy-fingered'' Dawn. Furthermore, epic poems feature stereotyped formulas for going to bed and getting up, putting on and taking off armour, sacrificing and feasting, and launching and beaching ships \cite{feyerabend1975}. 

\begin{quote}
    Scholars [showed] how the formulas varied with great subtlety and effect in relation to the specific narrative contexts in which they appeared. Often these variations were among different traditional elements, not between a traditional formula and a unique expression, suggesting that oral aesthetics consisted of a skillful use of traditional elements rather than the invention of new material. \cite{Beck2008}
\end{quote}

I argue then that the aforementioned endless cycle in which a problem description is constantly reinvented in every new publication has a detrimental effect on memory, and on quality of published material, since scholars must familiarise with endless modelling conventions and choices. It also makes it more difficult to carry out a literature survey and follow a given research thread across multiple publications, which may leverage different --- and in extreme cases even conflicting --- problem definitions. The approach here proposed, by promoting standardisation of problem definitions, is likely to ease these problems.

\section{Curatio in practice}\label{sec:practice}

After outlining the general research dissemination framework advocated in this manuscript, we now turn our attention to implementation, and discuss how this framework can be operationalised in practice. For the sake of convenience, we shall illustrate a possible implementation that blends \LaTeX{} and GitHub; for this purpose, a new \LaTeX{} package defining the environment \texttt{copyleft} has been developed; this is presented in Appendix \ref{sec:appendix}. 

\newpage
The environment is structured as follows,
\begin{verbatim}
\begin{copyleft}{Author}{Title}{Source}{License}
    Copyleft material.
\end{copyleft}
\end{verbatim}
Environment \texttt{copyleft} surrounds a block of copyleft material and, in line with copyleft attribution best practices,\footnote{\url{https://creativecommons.org/use-remix/attribution/}} accepts four input parameters: the author, the title of the work, the source of the work, and the licence under which the material is released.

Assume I have recently developed a new definition of the Knapsack Problem, and I have made it available in a GitHub repository under a Creative Commons Attribution 2.0 Generic (CC BY 2.0).\footnote{\url{https://creativecommons.org/licenses/by/2.0/}} The repository will contain a \LaTeX{} file \texttt{knapsack.tex}, which contains the \LaTeX{} code illustrated in Listing \ref{lst_knapsack}.\\

\begin{lstlisting}[frame=single,caption={knapsack.tex; observe how the Knapsack Problem definition is surrounded by the \texttt{copyleft} environment, and therefore annotated with information on author, title, source, and license, in line with copyleft attribution best practices.},label=lst_knapsack,captionpos=b,basicstyle=\ttfamily\scriptsize]
\begin{copyleft}
{Roberto Rossi}                                        % Author
{Knapsack Problem}                                     % Title
{\url{https://github.com/.../knapsack.tex}}            % Source
{Creative Commons Attribution 2.0 Generic (CC BY 2.0)} % License

Given a set of $n$ items numbered from 1 up to $n$, 
each with a weight $w_i$ and a value $v_i$, 
along with a maximum weight capacity $W$, the problem 
is to maximize the knapsack value, subject to the 
knapsack's capacity constraint, that is 
\[
\begin{array}{ll@{}ll}
\mathrm{max}        & \displaystyle\sum\limits_{i=1}^{n} v_{i}&x_{i} &\\
\mathrm{subject~to} & \displaystyle\sum\limits_{i=1}^{n} w_{i}&x_{j} \leq W\\
                    &                                        
                    &x_{i} \in \{0,1\}, &i=1 ,\dots, n.
\end{array}
\]
where $x_{i}$ represents the number of instances of item $i$ 
to include in the knapsack.

\end{copyleft}
\end{lstlisting}

The latex material in Listing \ref{lst_knapsack} can be easily embedded into any other latex document as follows,
\begin{verbatim}
\input{knapsack.tex}
\end{verbatim}
and this leads to the following result.
\begin{quote}
    \input{knapsack.tex}
\end{quote}

Assume now that an enhanced problem definition has been recently released by John Doe, and it has been made available as \texttt{better\_knapsack.tex} in a new GitHub repository, again under a Creative Commons Attribution 2.0 Generic (CC BY 2.0) license. We can embed this enhanced problem definition in our manuscript via the command
\begin{verbatim}
\input{better_knapsack.tex}
\end{verbatim}
and this leads to the following result.
\begin{quote}
\input{better_knapsack.tex}
\end{quote}

The use of package \texttt{copyleft} ensures that the original \LaTeX{} source is duly annotated in line with copyleft attribution best practices. Moreover, these annotations (author, title, source, and license information) are gathered and then compiled into a list of copyleft credits that can be printed after the usual list of references of an article via the command \texttt{\textbackslash printcopyleft}.\footnote{The list of copyleft credits for the present work is printed at the end of this document.} Note that improved versions of this package may compile and display these attributions in different forms, depending on the needs of the publication outlet.

\section{Conclusions}\label{sec:conclusions}

To conclude, in this paper I advocate a departure from existing research publication practices, which mainly rely upon copyrighted work for dissemination; and a move towards a new, more balanced, blend of copyright and copyleft, for dissemination of research results. 
Arguably, such a blend may facilitate the evolution and emergence of improved problem descriptions, whilst at the same time preserving author's rights, and easing researchers' work.
Finally, I have illustrated a possible strategy to operationalise this framework; this strategy leverages a newly developed \LaTeX{} package (\texttt{copyleft.sty}), and open repositories such as GitHub to distribute copyleft material.

\newpage

\appendix

\section{The \texttt{copyleft} \LaTeX{} package}\label{sec:appendix}

The \texttt{copyleft} \LaTeX{} package, presented in Listing \ref{lst:copyleft}, can be used in a latex manuscript by leveraging command
\begin{verbatim}
\usepackage{copyleft}
\end{verbatim}

\begin{lstlisting}[frame=single,caption=copyleft.sty,label=lst:copyleft,captionpos=b,firstline=17,lastline=52]
\NeedsTeXFormat{LaTeX2e}[1994/06/01]
\ProvidesPackage{copyleft}[2021/08/07 Copyleft Package]

\RequirePackage{etoolbox}

\def\copyleftlist{}%
\listadd{\copyleftlist}{}% Initialize list

\newrobustcmd{\myexpandingcommand}[1]{%
\listgadd{\copyleftlist}{#1}% Add an element
}%

% Macro showing the current list element%
\newrobustcmd{\showlist}[1]{%
#1%
}%

\newcommand\copyleftsource{}
\newcommand\copyleftlicence{}
\newenvironment{copyleft}[4]{
% Four arguments: title, author, source, license
% https://creativecommons.org/use-remix/attribution/
\expandafter\myexpandingcommand
\expandafter{#1. #2. #3. #4.}
}{}

\newcommand{\printcopyleft}{%
  \section*{Credits}
  This manuscript embeds copyleft material 
  from the following sources.
  \forlistloop{\showlist\\\\}{\copyleftlist}
}

\endinput
\end{lstlisting}

\newpage

\bibliographystyle{plain}
\bibliography{references}

\printcopyleft
\end{document}

%% file: knapsack.tex
\begin{copyleft}
{Roberto Rossi}                                        
{Knapsack Problem}                                     
{\url{https://github.com/.../knapsack.tex}}            
{Creative Commons Attribution 2.0 Generic (CC BY 2.0)} 

Given a set of $n$ items numbered from 1 up to $n$, 
each with a weight $w_i$ and a value $v_i$, 
along with a maximum weight capacity $W$, the problem 
is to maximize the knapsack value, subject to the 
knapsack's capacity constraint, that is 
\[
\begin{array}{ll@{}ll}
\mathrm{max}        & \displaystyle\sum\limits_{i=1}^{n} v_{i}&x_{i} &\\
\mathrm{subject~to} & \displaystyle\sum\limits_{i=1}^{n} w_{i}&x_{j} \leq W\\
                    &                                        
                    &x_{i} \in \{0,1\}, &i=1 ,\dots, n.
\end{array}
\]
where $x_{i}$ represents the number of instances of item $i$ 
to include in the knapsack.

\end{copyleft}

%% file: better_knapsack.tex
\begin{copyleft}
{John Doe}                                              
{A Better Knapsack Problem}                             
{\url{https://github.com/.../better_knapsack.tex}}      
{Creative Commons Attribution 2.0 Generic (CC BY 2.0)}  

Given a set of $n$ items numbered from 1 up to $n$, 
each with a weight $w_i$ and a value $v_i$, 
along with a maximum weight capacity $W$, the 
Knapsack Problem (KP) is to maximise the value of a 
knapsack (i.e. a selection of items), subject to the 
constraint that items picked must fit into its capacity.
The problem can be formulated mathematically as follows,
\[
\begin{array}{ll@{}ll}
\mathrm{max}        & \displaystyle\sum\limits_{i=1}^{n} v_{i}
                                                    &x_{i} &\\
\mathrm{subject~to} & \displaystyle\sum\limits_{i=1}^{n} w_{i}
                                               &x_{j} \leq W\\
                    &                                        
                            &x_{i} \in \{0,1\}, &i=1 ,\dots, n;
\end{array}
\]
where $x_{i}$ represents the number of instances of item $i$ 
to include in the knapsack.

\end{copyleft}